# Gerrymandering and geographic polarization have reduced electoral competition

**Ethan Jasny**[†]
Harvard College

**Christopher T. Kenny**[†]
Data-Driven Social Sciences
Princeton University
ctkenny@princeton.edu

**Cory McCartan**[†]
Department of Statistics
Pennsylvania State University
mccartan@psu.edu

**Tyler Simko**[†]
Department of Political Science
University of Michigan
tsimko@umich.edu

**Melissa Wu**[†]
Harvard College

**Michael Y. Zhao**[†]
Harvard College

**Aneetej Arora**
Department of Computer Science and Engineering
Ohio State University
arora.292@osu.edu

**Emma Ebowe**
Department of Government
Department of Africana Studies
College of William & Mary
evebowe@wm.edu

**Philip O'Sullivan**
Department of Statistics
Harvard University
posullivan@g.harvard.edu

**Taran Samarth**
Department of Political Science
Yale University
taran.samarth@yale.edu

**Kosuke Imai**[†]
Department of Government
Department of Statistics
Harvard University
imai@harvard.edu

August 21, 2025

## Abstract

Changes in political geography and electoral district boundaries shape representation in the United States Congress. To disentangle the effects of geography and gerrymandering, we generate a large ensemble of alternative redistricting plans that follow each state's legal criteria. Comparing enacted plans to these simulations reveals partisan bias, while changes in the simulated plans over time identify shifts in political geography. Our analysis shows that geographic polarization has intensified between 2010 and 2020: Republicans improved their standing in rural and rural-suburban areas, while Democrats further gained in urban districts. These shifts offset nationally, reducing the Republican geographic advantage from 14 to 10 seats. Additionally, pro-Democratic gerrymandering in 2020 counteracted earlier Republican efforts, reducing the GOP redistricting advantage by two seats. In total, the pro-Republican bias declined from 16 to 10 seats. Crucially, shifts in political geography and gerrymandering reduced the number of highly competitive districts by over 25%, with geographic polarization driving most of the decline.

**Keywords**: redistricting • gerrymandering • geography • partisan sorting

---

# 1. Introduction

In an attempt to consolidate partisan advantage in the upcoming midterm election, the Texas Republican-controlled legislature—encouraged by President Trump—approved a new congressional district map designed to potentially yield five additional seats for their party. Now, other states, including California, Florida, and New York, are considering similar mid-decade redistricting, not because of legal mandates but as calculated partisan responses.

These developments mark an escalation in a long-simmering redistricting war. In 2010, the Republican State Leadership Committee launched the Redistricting Majority Project (REDMAP), leveraging sophisticated mapping technologies to redraw congressional districts in swing states. Democrats responded in 2016 with the formation of the National Democratic Redistricting Committee (NDRC), aiming to counterbalance these efforts. By the 2020 redistricting cycle, both parties were engaged in partisan gerrymandering across many states (Kenny et al. 2023).

What drives the recent intensification of gerrymandering by both parties? Scholars have long observed that political geography in many states tends to advantage Republicans electorally (Chen and Cottrell 2016; Tamas 2019). Because Democratic voters are often concentrated in urban areas, district maps can exhibit "unintentional gerrymandering"—a pattern where a few overwhelmingly Democratic districts emerge, rather than a more even distribution of Democratic voters across multiple competitive districts (Chen and Rodden 2013; Mettler and Brown 2022; Rodden 2019).

However, more recent analyses suggest that this structural advantage for Republicans may have diminished or even disappeared entirely over the past decade (Stephanopoulos, McGhee, and Warshaw 2025). Since these broad shifts in political geography have occurred while partisan gerrymandering has increased, it remains unclear how these factors have together shaped the electoral environment since 2010.

We present the first comprehensive analysis of how both partisan geography and gerrymandering have changed across the 2010 and 2020 cycles. In particular, we disentangle the effects of change in political geography from those of partisan gerrymandering. To do so, we employ a state-of-the-art simulation algorithm (McCartan and Imai 2023) to generate large ensembles of alternative redistricting plans that adhere to each state's specific redistricting rules in both the 2020 and 2010 redistricting cycles (McCartan et al. 2022). Our simulations improve upon previous work (Chen and Cottrell 2016) by expanding temporal coverage and incorporating key redistricting criteria such as the avoidance of county splits and compliance with the Voting Rights Act (VRA).

These simulated plans serve as a non-partisan baseline, allowing us to quantify the extent of partisan gerrymandering by comparing enacted maps against this benchmark. They also allow us to trace how political geography within each state has evolved across redistricting cycles over the past decade. While prior work (Warshaw, McGhee, and Migurski 2022) has documented changes in gerrymandering over time, it lacked such a baseline and therefore could not disentangle shifts in political geography from changes due to map drawing.

Our nonpartisan simulation analysis reveals a marked increase in urban-rural political polarization at the district level since 2010. The most rural congressional districts have become more Republican, while many urban districts have grown more Democratic. In the top 25% most urban districts, the average Democratic win probability rose from roughly 80% in 2010 to 88% in 2020. Conversely, in the top 25% most rural districts, the average Democratic win probability fell from 20% to 12%. Although these opposing geographic trends largely offset each other at the national level, Democrats gained a slight net advantage due to improvements in competitive suburban districts. As a result, the Republican geographic advantage declined from roughly 14 seats in 2010 to 9 seats in 2020.

Like geographic polarization, partisan gerrymandering has also intensified over the past decade. In the 2010 redistricting process, we find 11 states deviated at a statistically detectable level from our nonpartisan simulations—eight of which favored Republicans. By 2020, however, Democrats made redistricting gains to shrink this Republican advantage. Democrats mitigated GOP advantages in Rust Belt states like Ohio and instituted new pro-Democratic maps in states such as Illinois. Overall, the net Republican advantage from partisan gerrymandering declined by two seats between 2010 and 2020.

Notably, states with the largest Republican shifts in political geography also show the most substantial countervailing gerrymandering in favor of Democrats. Texas, where political geography shifted most strongly toward Democrats, experienced the most aggressive Republican gerrymander in 2020, amounting to around two additional GOP-leaning seats. The recently enacted gerrymander will likely increase this large existing advantage even further. In this way, partisan redistricting efforts largely served to blunt the electoral impact of evolving geographic patterns. In broad terms, although enacted plans often distorted partisan balance within individual states, the aggregate effect across states was to largely preserve the existing national partisan balance.

In total, the Republican structural advantage in the U.S. House declined modestly from about 16 seats in 2010 to 10 seats in 2020. Of this six-seat shift, we attribute four seats to changes in geographic polarization and two seats to the impact of partisan gerrymandering. Put differently, even if Democrats had won 50% of the national vote in 2020, they would have fallen about 10 seats short of a House majority.

Most importantly, both geographic polarization and gerrymandering has substantially reduced electoral competition. Geographic polarization alone reduced the number of highly competitive House districts—those with an expected margin of victory under five percentage points—from approximately 67 in 2010 to just 50 in 2020. Partisan gerrymandering exacerbated this trend; relative to the nonpartisan baseline, redistricting further reduced the number of competitive seats from 67 to 47 in 2010 and from 50 to 34 in 2020.

This erosion of electoral competition poses a major threat to the health of American democracy (Issacharoff 2002). As the number of competitive seats dwindles, control of the House increasingly hinges on a narrow set of swing districts. Consequently, the chamber becomes less responsive to shifts in national public opinion. Moreover, in the absence of meaningful general election challenges, representatives in safe seats face little incentive to appeal to the political center, encouraging ideological extremism and deepening partisan polarization (Carson et al. 2007).

## 2. Changes in Geographic Polarization

America's modern political geography is defined by an urban-rural divide: cities are overwhelmingly Democratic and rural areas are strongly Republican (Rodden 2019). This trend has become entrenched, following decades of geographic realignment along partisan lines (Brown and Mettler 2024). However, it remains an open question how changes in political geography aggregate to legislative districts over time. This question poses a difficult measurement challenge, because districts can be drawn intentionally to favor political parties. Our approach allows us to decompose changes in geographic polarization from this gerrymandering.

We measure the urbanity of all districts in each of our simulated plans from the 2010 and 2020 cycles. Using the 2020 Census' designation of rural and urban areas, we calculate the proportion of voters in each simulated district who belong to urban census blocks as defined by the Census Bureau. By comparing the simulated plans between 2010 and 2020, we are able to quantify the degree to which geographic polarization has advanced without being confounded by partisan gerrymandering.

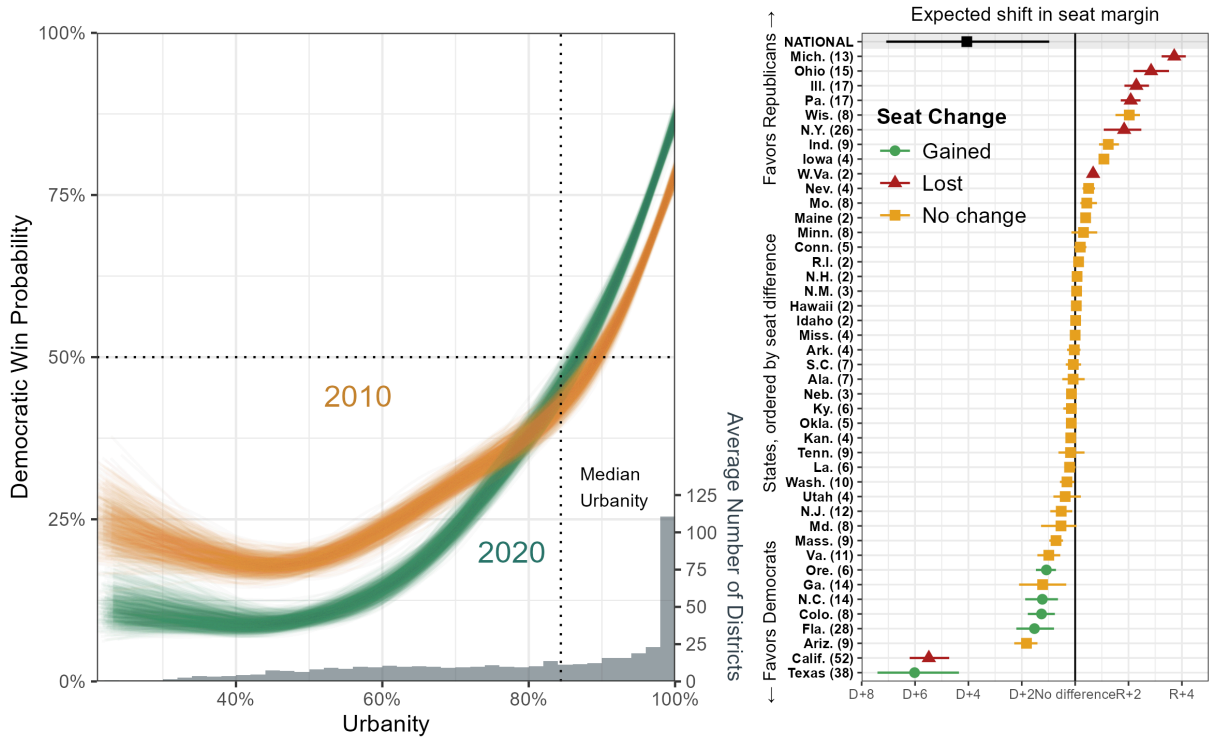

**Figure 1:** The left panel shows how the average Democratic win probability varies with the urbanity of simulated districts. Each line represents the estimates based on a simulated plan. A random sample of 1,000 simulated plans is used for both 2010 and 2020. The gray histogram plots the distribution of urban districts, averaging across 2010 and 2020. The dotted vertical line represents the urbanity of the median simulated district, which is roughly 84% urban. The right panel shows estimated change in expected seat margin from 2010 to 2020. Estimates show the expected Republican seats under each nonpartisan baseline plan in 2010 subtracted from the expected Republican seats under each nonpartisan baseline plan in 2020. The horizontal bars show 95% intervals of the simulated plans.

The left plot of Figure 1 shows urban-rural polarization has grown considerably over the past decade. The yellow and green curves plot the relationship between urbanity and the average probability of Democratic victory, with each line representing the estimates based on a simulated plan. The gray histogram plots the distribution of urbanity across districts, where the median district is around 84% urban.

Between 2010 and 2020, urban districts—which already leaned heavily Democratic—became even more Democratic. The average Democratic win probability increased from 80% in 2010 to 88% in 2020 in the top 25% most urban districts. Meanwhile, the most rural districts became even more Republican. The average Democratic win probability dropped from 20% to 12% in the top 25% most rural districts. As described in the Materials and Methods section, these win probabilities, and all of the partisan outcome estimates presented here, are based on a probabilistic election model that averages over election-to-election variation from a national baseline election.

As shown in the right panel of Figure 1, these shifts in win probability largely offset each other nationally, with Democrats gaining a little more than four seats in 2020 compared to 2010. While this analysis of the 2010–2020 period reveals a modest gain for Democrats, they remain disadvantaged overall.

Most of the effects of geographic polarization happen in deeply urban or deeply rural districts that tend to be already safe for one of the parties. In a highly geographically polarized environment, what matters instead are the relatively few competitive districts hovering around median urbanity. In these suburban and exurban districts, we find slight gains for Democrats. Still, Republicans remain advantaged in the overall House geography. Democratic gains merely attenuated the net geographic bias from around 14.3 seats in 2010 to 9.8 seats in 2020.

The polarization by urbanity occurred simultaneously with larger regional shifts in party strength. The right panel of Figure 1 also shows the breakdown of the geographic shifts at the state level. Republicans made major gains in the Midwest and the Rust Belt (Michigan, Ohio, Illinois, Pennsylvania, Wisconsin). The congressional geography similarly improved for Democrats in some Sun Belt states (e.g., Texas, California, Arizona). As these changes are in opposite directions, the *net* change remains small.

## 3. Changes in Partisan Gerrymandering

We find that partisan gerrymandering has intensified over the past decade. In 2010, enacted plans in 11 states deviated at a statistically detectable level from our nonpartisan simulations—three favoring Democrats and eight favoring Republicans (see Figure A4). By 2020, the number of states with statistically significant gerrymanders had risen to 15, with seven advantaging Democrats and eight advantaging Republicans (see Figure A3).

Our analysis shows that partisan gerrymandering has often worked to offset shifts in political geography. The left panel of Figure 2 plots the change in partisan bias of enacted plans from 2010 to 2020. In several Midwestern and Rust Belt states—such as Michigan, Illinois, and Pennsylvania where political geography shifted substantially toward Republicans (see the right panel of Figure 1), enacted maps produced notable changes in favor of Democratic party. Conversely, Texas, which experienced the largest baseline shift in political geography toward Democrats, saw the single largest increase in pro-Republican partisan bias, equivalent to roughly two seats. In effect, gerrymandering slowed the impact of changing political geography. Overall, Democrats made redistricting gains relative to Republicans, narrowing the GOP's gerrymandering advantage by about two seats.

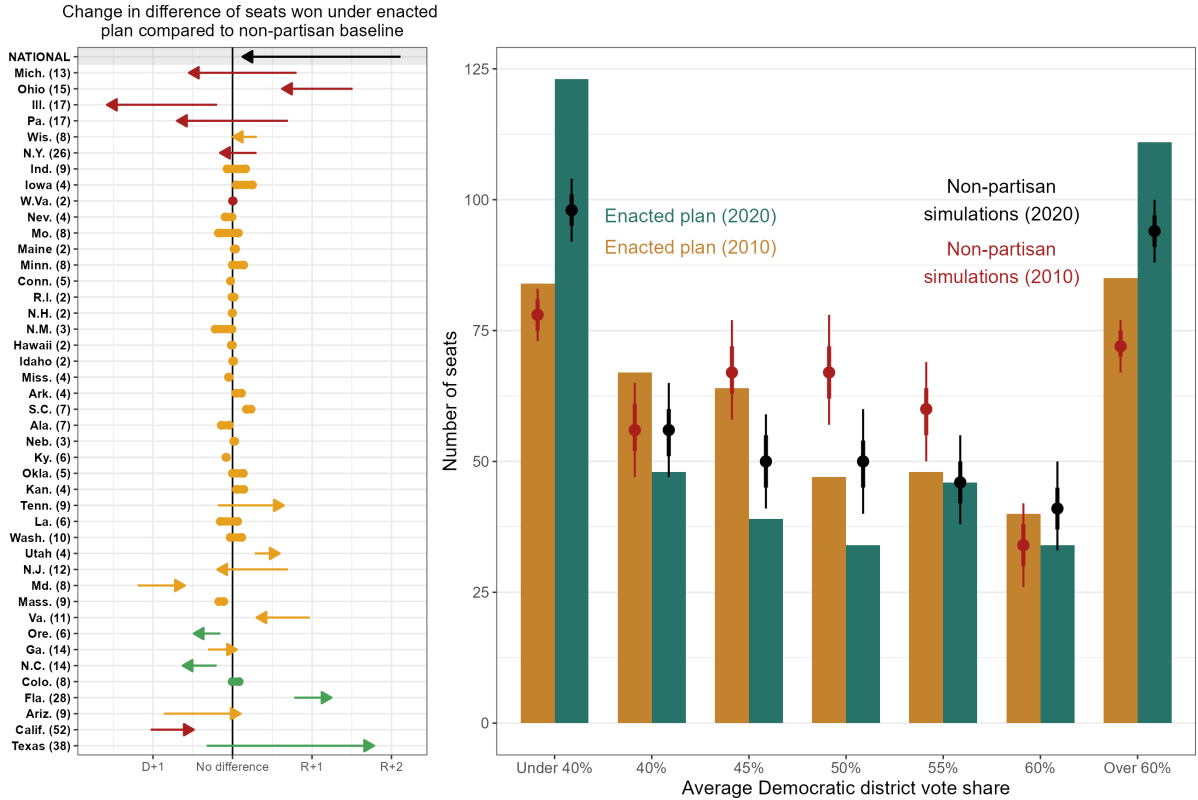

**Figure 2:** The histogram in the right panel shows the distribution of expected district vote shares under the enacted districts for 2010 (yellow) and 2020 (green). The overlaid 66% and 95% confidence intervals show the range of the same quantity under our nonpartisan simulated baseline for 2010 (red) and 2020 (black). The left panel shows the change in partisan gerrymandering between 2010 and 2020. Arrows start at the difference between the enacted plan and average simulated plan in 2010 and end at the difference between the enacted plan and average simulated plan in 2020. Arrows that point to the left experienced an increase in pro-Democratic gerrymandering, while arrows that point to the right experienced an increase in pro-Republican gerrymandering. Colors are determined by whether the state lost or gained seats in the 2020 apportionment cycle.

The magnitude of overall redistricting bias in both 2010 and 2020 depends on the assumed national electoral environment. Our main analysis uses a 50–50 national national environment, but alternative baselines are also possible. As shown in Appendix Section Section C, adopting a 51.5% Democratic baseline —the Democrats' average national two-party vote share in 2016 and 2020—still produces a Republican gerrymandering advantage, consistent with the findings of Kenny et al. (2023). Yet, regardless of the baseline, political geography remains the greater source of Republican advantage, contributing about 9.8 seats in 2020, more than gerrymandering itself.

## 4. Decline in Electoral Competition
Between 2010 and 2020, the competitiveness of enacted congressional plans declined substantially. As shown in the center two bars of Figure 2 (right), the number of competitive seats—those where the Republican and Democratic two-party vote shares are within 5 percentage points—fell by more than 25 percent, dropping from 47 to just 34. Using the baseline simulated plans, we can again separate this decline into the effects of changing geography and of partisan line-drawing.

Both geographic polarization and partisan gerrymandering have contributed to declining electoral competition over the past decade. We estimate that partisan gerrymandering alone eliminated about 20 competitive seats in 2010 and 16 in 2020, relative to the simulated baseline. In both cycles, this represented roughly a 30 percent reduction in competitive districts. Figure 2 (right) demonstrates that the enacted plans (solid bars) consistently contain far fewer competitive seats than the simulations (confidence intervals with point estimates) and, conversely, many more safe seats. For instance, in 2020, the enacted plans contained 42 additional safe seats—those expected to be decided by at least 20 percentage points in the two-party vote—than the simulations predicted.

Although partisan gerrymandering depressed competitiveness in both decades, the change over time is driven more strongly by geographic polarization. Between 2010 and 2020, the simulated plans lost about 17 competitive seats, compared with a loss of roughly 13 in the enacted plans. As rural areas became more Republican and urban areas more Democratic, regions with relatively balanced partisan populations became scarcer. These results suggest that even if map-drawers had sought to design more competitive districts in 2020, doing so would have been harder than in 2010.

## 5. Discussion

In anticipation of the 2010 redistricting cycle, Republicans invested millions of dollars in efforts to strategically draw the lines that determine how American democracy operates. Our simulation analysis shows that this coordinated effort resulted in a net Republican seat advantage. Depending on the choice of electoral baseline, this advantage due to gerrymandering alone gained Republicans several additional seats above their existing geographic advantage.

While the 2010 cycle was hailed as a Republican win, as we show, this was largely due to their geographic advantage. Between 2010 and 2020, the overall pro-Republican geographic bias of the House of Representatives was largely unchanged, with Democrats needing a bit more than 51% of the vote to capture a majority in the chamber after both cycles. But, Democrats have made some modest gains, reducing the Republican geographic advantage by around four seats.

On the redistricting front, Democrats joined the fray in 2020, diminishing the Republican gerrymandering advantage to nearly zero, on average. Notably, widespread Democratic and Republican gerrymanders partially stalled the shifts in political geography measured by the simulations: enacted plans in 2020 tended to shield the status quo from changing geography.

Of utmost importance, geographic polarization and gerrymandering combined to reduce the competitiveness of congressional elections. From 2010 to 2020, the number of seats expected to be decided by at most five percentage points was substantially reduced. While partisan bias obviously affects the balance of power, a decline in competitiveness also has serious consequences. With fewer closely contested races, general election votes matter less, limiting voters' ability to hold their representatives accountable. In uncompetitive districts, the primary rather than the general election becomes the decisive contest. As a result, members of Congress are often more accountable to primary electorates than to the broader public, a dynamic that may further reduce incentives for cross-party compromise (Anderson, Butler, and Harbridge-Yong 2020; Carson et al. 2007).

Given ongoing attempts to redistrict states mid-cycle, it is crucial to distinguish structural shifts in political geography from the strategic choices of partisan map-drawers. Disentangling these forces allows us to better understand and counteract their effects on electoral competitiveness. Political manipulation in redistricting can be curbed through nonpartisan commissions (McCartan et al. 2025), whereas addressing the consequences of geographic polarization may require more fundamental changes in electoral institutions, such as adopting proportional representation or multi-member districts.

## A. Materials and Methods

State and federal rules constrain how congressional districts are drawn. Federal constitutional constraints require that districts be approximately equal in population. Districts must also comply with the Voting Rights Act of 1965, which enforces protections for racial minorities. Historically, contiguity was federally required and has become standard in all states, some enacting their own contiguity requirements. Most states, at least by norm, also expect districts to be geographically compact and avoid egregiously splitting counties and municipalities. These federal requirements and traditional criteria are supplemented by state-specific criteria, which vary widely, but influence how plans may be drawn.

We generate simulated plans that adhere to each state's specific redistricting rules using the Sequential Monte Carlo (SMC) redistricting algorithm of McCartan and Imai (2023). Simulations using SMC and following each state's criteria for the 2020 cycle are described in detail in McCartan et al. (2022). We follow the same procedure to generate simulations for the 2010 cycle and introduce them here. For the 2010 cycle, we use the 2010 Census data and the corresponding state-specific redistricting criteria in place at the time. For each state and cycle, we have 5,000 simulated redistricting plans that closely follow state's legal criteria. The full code to generate the simulations is available for both cycles at https://github.com/alarm-redist/fifty-states. McCartan et al. (2022) discuss the limitations of the simulated districts, which stem in large part from the inherent imprecision of state and federal criteria.

There are three kinds of electoral summaries of each enacted and simulated district or plan that we use in this paper:

1. Average Democratic vote share in a district (Fig 2, right)
2. Average Democratic win probability in a district (Fig 1, left)
3. Average number of seats won by Democrats in a plan (Fig 1, right and Fig 2, left)

All of these numbers are based on a probabilistic election model that averages over election-to-election variation.

The model we use is a version of the stochastic uniform partisan swing model of Gelman and King (1994) and is described fully in Kenny et al. (2023). The model is based on a precinct-level baseline vote, which is then shifted in each election by a national swing, shared across districts, and a district-specific swing that captures local variation. Here, the precinct-level baseline vote is the two-party presidential vote share for elections in the five years preceding redistricting. Thus we use the 2008 presidential vote share for the 2010 cycle and the average of the 2016 and 2020 presidential vote shares for the 2020 cycle.

Comparing electoral baselines across time is necessary to allow for changing geographic patterns of voters and population changes over time. However, this introduces a potential issue, where the national two party vote share also shifts between elections. Specifically, the vote share for Obama in 2008 was about 53.5% of the two-party vote, while an average of Clinton in 2016 and Biden in 2020 received about 51.5% on average. Comparing these patterns directly would necessarily lead to a change in the partisan bias of the enacted plans, even if the underlying geography and gerrymandering remain unchanged.

To avoid comparing different electoral environments, we shift the data to a 50% national vote share for both cycles on the logit scale. A 50-50 partisan split between the two major parties allows for a natural comparison of the partisan bias of enacted plans and the simulated plans. For consistency with Kenny et al. (2023), we also present results under a 51.5% Democratic national vote share in Appendix Section C.

The average Democratic vote share in a district is then calculated directly by averaging the precinct-level vote shares in the district, weighted by the average precinct turnout. The average Democratic win probability in a district is calculated from the election model. Each baseline district vote share maps to an average win probability, based on the modeled random national and district-level swings. This model averaging is described in more detail in the appendices to Kenny et al. (2023). The average number of seats won by Democrats in a plan is then calculated by summing the average win probabilities across the districts in question.

## B. Raw Simulation and Gerrymandering Estimates

Estimates in the main text are focused on the change between decades. Here, we present estimates of the main simulation results and the partisan bias of the enacted plans for 2010 and 2020 separately. Each of these are shown under a 50% baseline.

Figure A1 shows the share of seats won by Democrats in each state under 5,000 simulated plans for the 2020 cycle, using the average of the 2016 and 2020 elections shifted to the 50% baseline. Figure A2 shows the corresponding results for the 2010 cycle, using the 2008 geographic distribution of votes shifted to the 50% baseline.

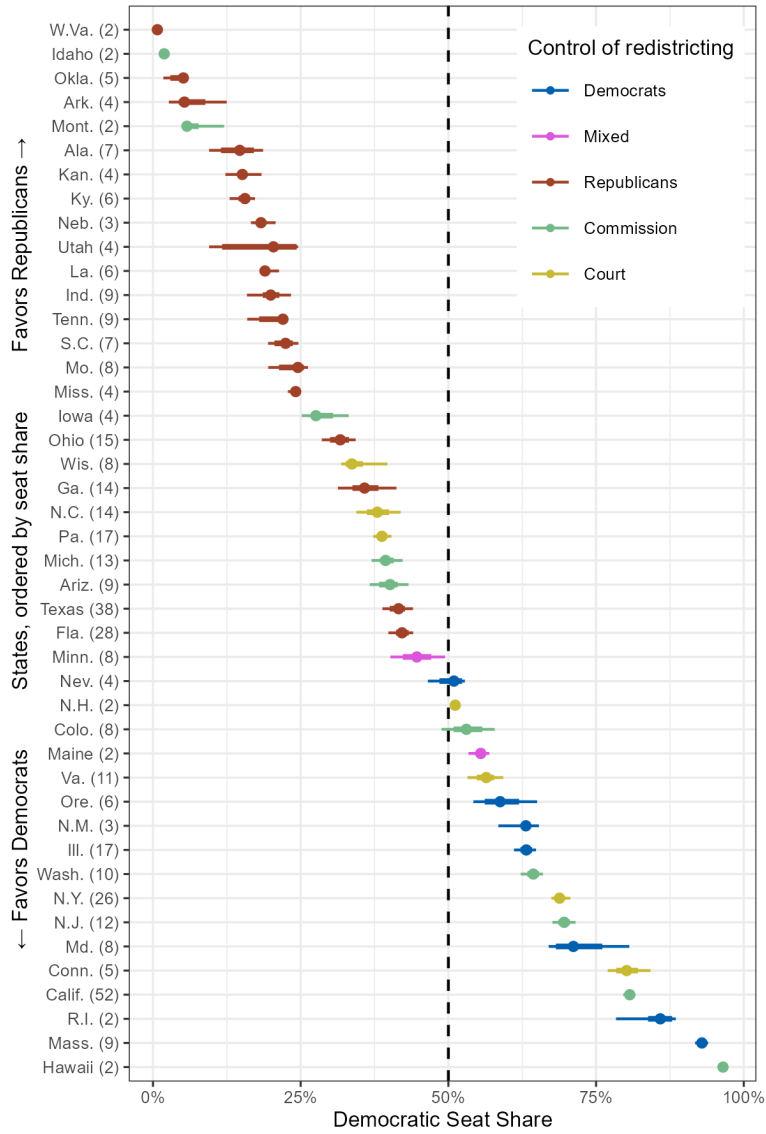

**Figure A1: Simulated plans for the 2020 cycle**. Each point shows the median Democratic share. The wide bar shows the middle 66% range of the simulations and the thin bar shows the 95% range. Points are colored by who controlled the redistricting process at the time that the plan was drawn. States near the top of the graph favor Republicans, while states near the bottom of the graph favor Democrats. Numbers in parentheses show the number of districts in the state.

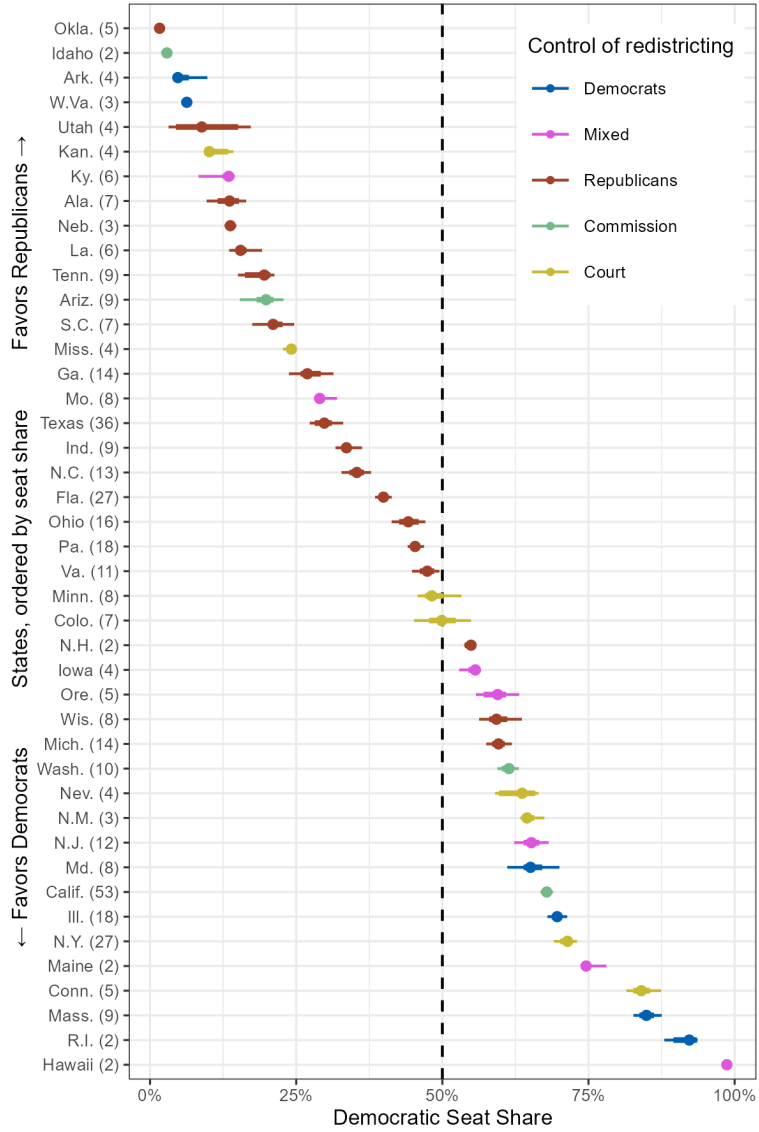

**Figure A2: Simulated plans for the 2010 cycle**. Each point shows the median Democratic share. The wide bar shows the middle 66% range of the simulations and the thin bar shows the 95% range. Points are colored by who controlled the redistricting process at the time that the plan was drawn. States near the top of the graph favor Republicans, while states near the bottom of the graph favor Democrats. Numbers in parentheses show the number of districts in the state.

Next, we provide our estimates of the redistricting bias by decade at the 50% electoral baseline. Figure A3 shows the redistricting bias of the enacted plans in 2020 at the state level. This figure replicates Figure 1 of Kenny et al. (2023), but with the electoral baseline shifted.

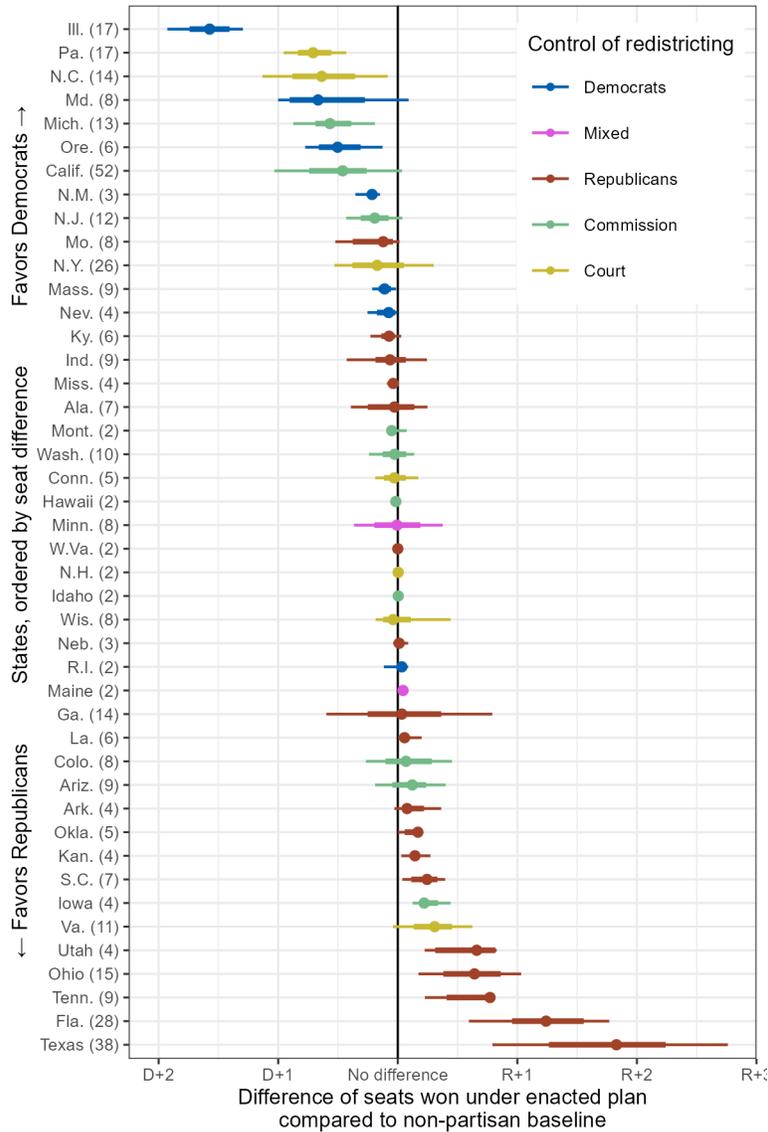

**Figure A3: Estimated redistricting bias by state**. Each point shows the median estimate of the difference between the simulated plans and the enacted plan for the 2020 cycle. Colors show which party had control of the redistricting institution which drew the plan. The wide bars show the middle 66% of the simulations and then thin bars show the 95% interval. The black line indicates no bias. Estimates to the left and right correspond to Democratic and Republican biases, respectively.

Figure A4 similarly shows the redistricting bias of the enacted plans in 2010 at the state level.

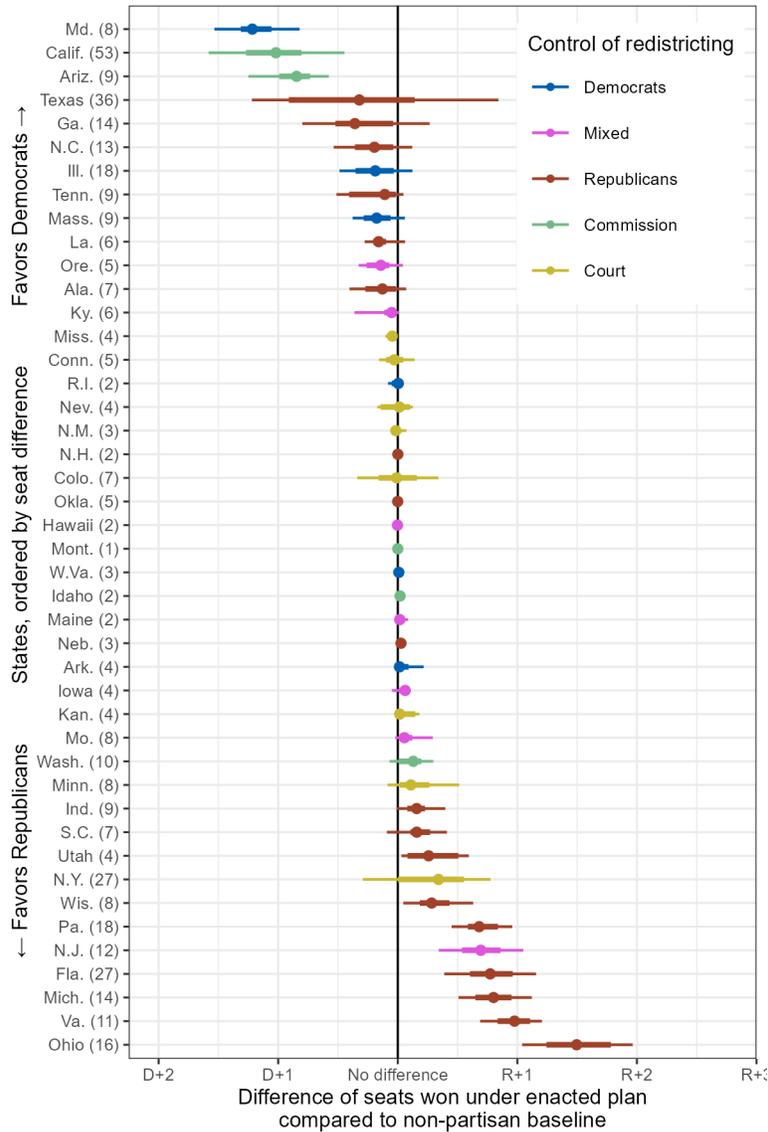

**Figure A4: Estimated redistricting bias by state**. Each point shows the median estimate of the difference between the simulated plans and the enacted plan for the 2010 cycle. Colors show which party had control of the redistricting institution which drew the plan. The wide bars show the middle 66% of the simulations and then thin bars show the 95% interval. The black line indicates no bias. Estimates to the left and right correspond to Democratic and Republican biases, respectively.

**Table A1: Point estimates for simulations and gerrymandering by cycle.** These present our estimates for the simulations, enacted plans, and their difference between them for each cycle. The difference of the simulations and enacted plans is labelled as the gerrymandering for that cycle. We also include 95% confidence intervals on their differences. The final row provides point estimates of the change across decades. These estimates correspond to the prior figures in this section.

| Year | Dem. Seats (Sims.) | Dem. Seats (Enac.) | Gerrymandering | (Lower) | (Upper) |
|------|------|------|------|------|------|
| 2010 | 203.68 | 201.56 | −2.11 | −4.20 | −0.15 |
| 2020 | 208.24 | 208.11 | −0.12 | −2.43 | 2.16 |
| Change | 4.56 | 6.55 | 1.99 | 0.97 | 7.08 |

## C. An Alternate National Baseline

In the main text, we set the national baseline to a 50-50 partisan split for both decades. Fixing this is important, as comparing different national baselines can lead to confusion about the source of changes in partisan bias. Particularly, a shift in the national baseline can lead to a change in the partisan bias of the enacted and simulated plans, even if the underlying geography and gerrymandering remain unchanged. For example, comparing the raw 2008 and 2020 elections would show a large shift in the geographic bias, as Obama received about 2 percentage points more of the two-party vote share in 2008 than Biden did in 2020.

There are many justifiable choices for the national baseline. Here, we shift the data to a 51.5% Democratic baseline, which is the average national two-party vote share for Democrats in 2016 and 2020. This baseline is identical to that used in Kenny et al. (2023). Below, we present the results of this alternative baseline.

Figure A5 (left) shows the relationship between urbanity and Democratic win probability under the 51.5% baseline. The results are extremely similar to those under the 50% baseline, but Democrats typically have a higher win probability, owing to the higher baseline.

Figure A5 (right) shows the change in simulated districts between 2010 and 2020 under a 51.5% baseline, which amounts to a D+0.3 change. The change in geography is almost entirely cancelled out by rightward shifts in the Rust Belt and the leftward shifts in the southwest.

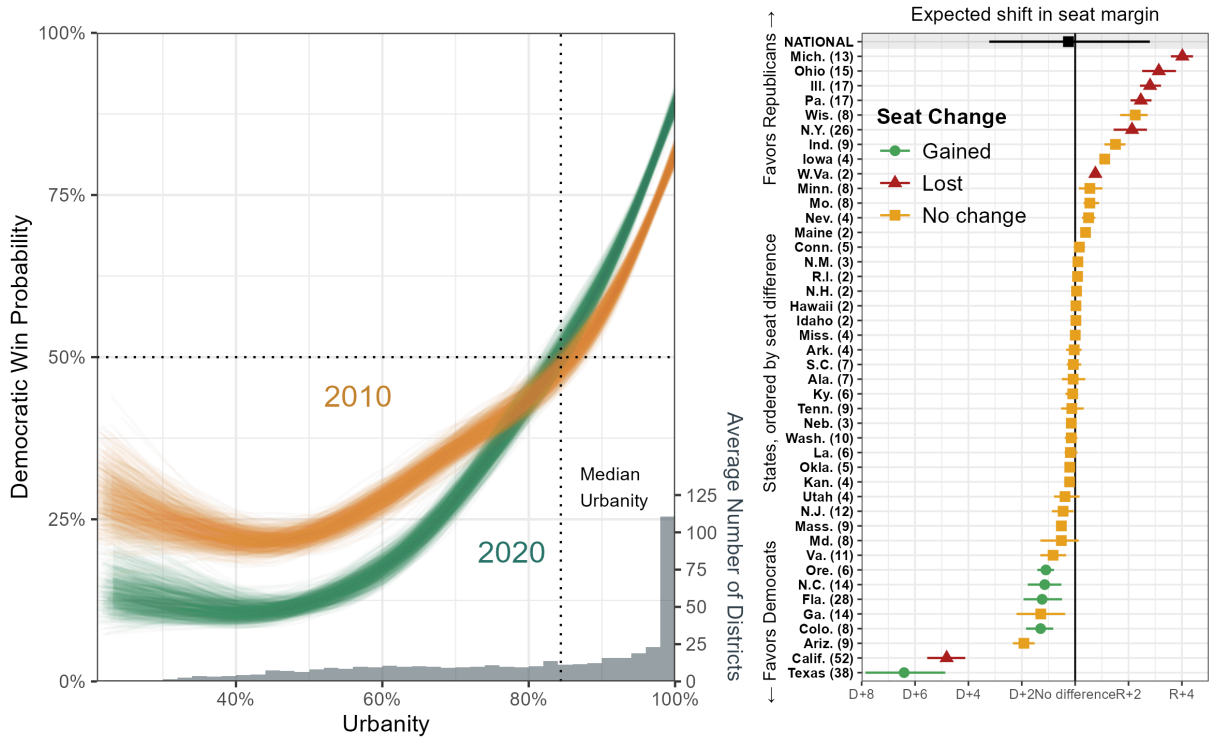

**Figure A5:** The left panel shows how the average Democratic win probability varies with the urbanity of simulated districts under a 51.5% baseline. Each line represents a simulation, with a generalized additive model used to fit the relationship between urbanity and Democratic win probability. A random sample of 1,000 simulations are plotted for both 2010 and 2020. The gray histogram plots the distribution of urban districts, averaging across 2010 and 2020. The dotted vertical line represents the urbanity of the median simulated district, which is roughly 84% urban. The right panel shows estimated change in expected seat margin from 2010 to 2020 under a 51.5% baseline. Estimates show the expected Republican seats under each nonpartisan baseline plan in 2010 subtracted from the expected Republican seats under each nonpartisan baseline plan in 2020. Bars to the right indicate that the simulated plans shift toward the Republicans between 2010 and 2020, while to the left indicate shift toward Democrats. The bars show 95% intervals of the simulated plans.

Figure A6 (left) displays the state-level changes in gerrymandering under the 51.5% baseline. As before, we see increases in Democratic gerrymandering in the Rust Belt and increases in Republican gerrymandering in the Sun Belt. The net change in gerrymandering is still about 2 Democratic seats, shifting from D+4.4 in 2010 to D+2.4 in 2020.

Figure A6 (right) repeats the scoring of competitive districts from Figure 2 (right), but under the 51.5% baseline. Here, we see qualitatively similar results, with enacted districts being less competitive than simulations on average. However, there is a smaller difference between the enacted and simulated plans in 2010.

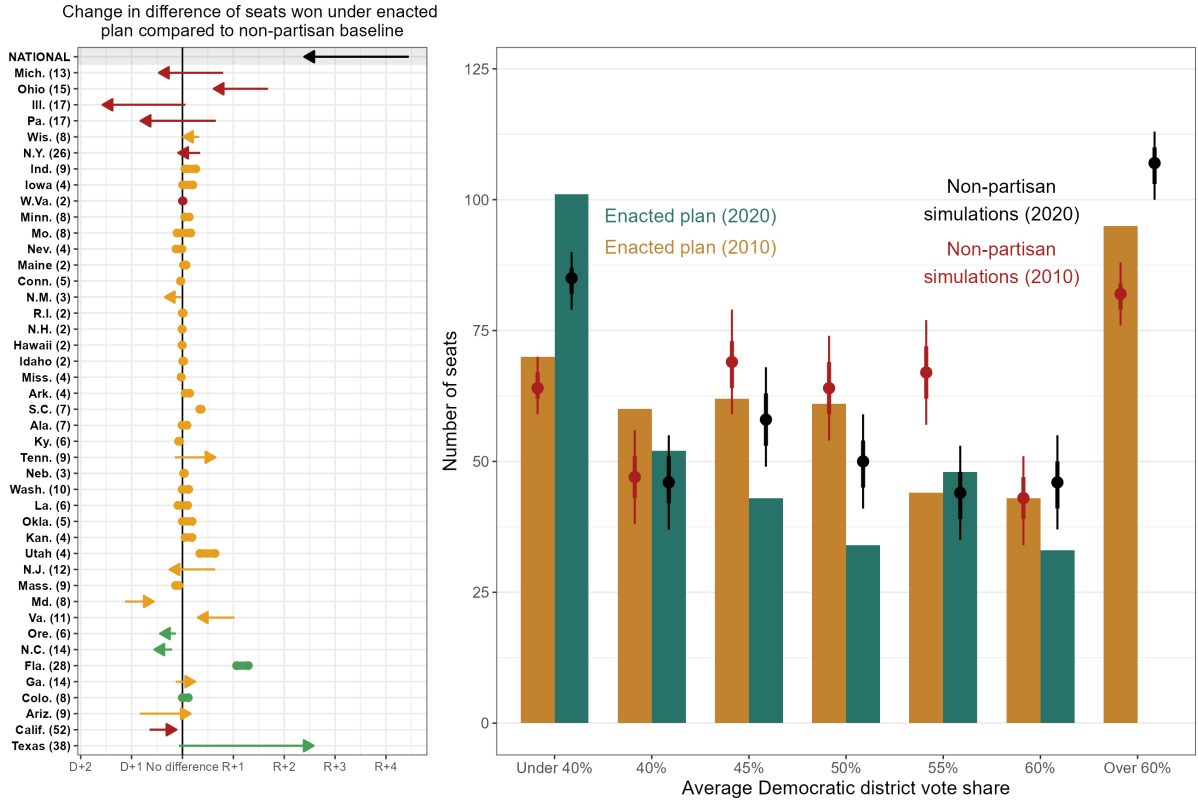

**Figure A6:** The histogram shows the distribution of expected district vote shares under the enacted districts for 2010 (yellow) and 2020 (green) under a 51.5% baseline. The overlaid 66% and 95% confidence intervals show the range of the same quantity under our nonpartisan simulated baseline for 2010 (red) and 2020 (black). The right panel shows the change in partisan gerrymandering between 2010 and 2020 under a 51.5% baseline. Arrows start at the difference between the enacted plan and average simulated plan in 2010 and end at the difference between the enacted plan and average simulated plan in 2020. Arrows that point to the left experienced an increase in pro-Democratic gerrymandering, while arrows that point to the right experienced an increase in pro-Republican gerrymandering. Colors and shapes are determined by whether the state lost or gained seats in the 2020 apportionment cycle.

Finally, we include a replicated table of key point estimates under the 51.5% baseline in Table A2.

**Table A2: Point estimates for simulations and gerrymandering by cycle under the 51.5% baseline.** These present our estimates for the simulations, enacted plans, and their difference between them for each cycle. The difference of the simulations and enacted plans is labelled as the gerrymandering for that cycle. We also include 95% confidence intervals on their differences. The final row provides point estimates of the change across decades.

| Year | Dem. Seats (Sims.) | Dem. Seats (Enac.) | Gerrymandering | (Lower) | (Upper) |
|------|------|------|------|------|------|
| 2010 | 223.37 | 218.92 | −4.45 | −6.51 | −2.50 |
| 2020 | 224.13 | 221.73 | −2.39 | −4.69 | −0.11 |
| Change | 0.75 | 2.81 | 2.05 | −2.80 | 3.22 |

# References

Anderson, Sarah E., Daniel M. Butler, and Laurel Harbridge-Yong. 2020. "Primary Voters as the Source of Punishment". In *Rejecting Compromise: Legislators' Fear of Primary Voters*, 55–81. Cambridge University Press.

Brown, Trevor E., and Suzanne Mettler. 2024. "Sequential Polarization: The Development of the Rural-Urban Political Divide, 1976–2020". *Perspectives on Politics* 22 (3): 630–58.

Carson, Jamie L., Michael H. Crespin, Charles J. Finocchiaro, and David W. Rohde. 2007. "Redistricting and Party Polarization in the U.S. House of Representatives". *American Politics Research* 35 (6): 878–904. https://doi.org/10.1177/1532673X07304263.

Chen, Jowei, and David Cottrell. 2016. "Evaluating Partisan Gains from Congressional Gerrymandering: Using Computer Simulations to Estimate the Effect of Gerrymandering in the U.S. House". *Electoral Studies*, no. 44 .

Chen, Jowei, and Jonathan Rodden. 2013. "Unintentional Gerrymandering: Political Geography and Electoral Bias in Legislatures". *Quarterly Journal of Political Science*, no. 8 , 239–69.

Gelman, Andrew, and Gary King. 1994. "A Unified Method of Evaluating Electoral Systems and Redistricting Plans". *American Journal of Political Science*, 514–54.

Issacharoff, Samuel. 2002. "Gerrymandering and Political Cartels". *Harvard Law Review* 116 (2): 593–648.

Kenny, Christopher T., Cory McCartan, Tyler Simko, Shiro Kuriwaki, and Kosuke Imai. 2023. "Widespread Partisan Gerrymandering Mostly Cancels Nationally, But Reduces Electoral Competition". *Proceedings of the National Academy of Sciences* 120 (25): e2217322120. https://doi.org/10.1073/pnas.2217322120.

McCartan, Cory, and Kosuke Imai. 2023. "Sequential Monte Carlo for Sampling Balanced and Compact Redistricting Plans". *Annals of Applied Statistics* 17 (4): 3300–3323.

McCartan, Cory, Christopher T. Kenny, Tyler Simko, Emma Ebowe, Michael Y. Zhao, and Kosuke Imai. 2025. "Redistricting Reforms Reduce Gerrymandering by Constraining Partisan Actors". *Arxiv Preprint 2407.11336.*

McCartan, Cory, Christopher T. Kenny, Tyler Simko, George Garcia III, Kevin Wang, Melissa Wu, Shiro Kuriwaki, and Kosuke Ima. 2022. "Simulated Redistricting Plans for the Analysis and Evaluation of Redistricting in the United States". *Scientific Data* 9 (689).

Mettler, Suzanne, and Trevor Brown. 2022. "The Growing Rural-Urban Political Divide and Democratic Vulnerability". *The ANNALS of the American Academy of Political and Social Science* 699 (1): 130–42.

Rodden, Jonathan. 2019. *Why Cities Lose: The Deep Roots of the Urban-Rural Political Divide.* Basic Books.

Stephanopoulos, Nicholas, Eric McGhee, and Christopher Warshaw. 2025. "The House's Republican Edge Is Gone. But the Gerrymander Lives.". *The Washington Post.*

Tamas, Bernard. 2019. "American Disproportionality: A Historical Analysis of Partisan Bias in Elections to the U.S. House of Representatives". *Election Law Journal* 18 (1).

Warshaw, Christopher, Eric McGhee, and Michal Migurski. 2022. "Districts for a New Decade—Partisan Outcomes and Racial Representation in the 2021–22 Redistricting Cycle". *Publius: The Journal of Federalism* 52 (3): 428–51. https://doi.org/10.1093/publius/pjac020.